# Assessment of Continuous-Time Transmission-Distribution-Interface Active and Reactive Flexibility for Flexible Distribution Networks

Shuo Yang, *Student Member, IEEE*, Zhengshuo Li, *Senior Member, IEEE,* and Ye Tian, *Student Member, IEEE*

*Abstract*—With the widespread use of power electronic devices, modern distribution networks are turning into flexible distribution networks (FDNs), which have enhanced active and reactive power flexibility at the transmission-distribution-interface (TDI). However, owing to the stochastics and volatility of distributed generation, the flexibility can change in real time and can hardly be accurately captured using conventional discrete-time (DT) assessment methods. This paper first proposes the notion of continuous-time (CT) TDI active and reactive flexibility and establishes its mathematical model. This model comprehensively considers the flexible devices in the FDN and the impact of uncertainty of photovoltaic power generation and load. In particular, a novel direction-factor-based metric is proposed to model CT-TDI PQ flexibility. Moreover, an efficient solution method is designed to address the difficulties in handling the infinite dimension of CT model and the complexity of bi-objectivity from assessing both active and reactive flexibility to be assessed. The solution successfully transforms the infinite dimensional optimization into a finite dimensional problem and effectively explores the PQ plane in a parallel pattern. Case studies show that the method can more effectively assess the real-time TDI flexibility of an FDN relative to conventional DT counterparts, and also reveals the impact of the relevant factors, such as penetrations of flexible devices and levels of uncertainty.

## I. Introduction

WITH the widespread use of advanced power electronic devices, modern distribution networks have become increasingly controllable and flexible, distinguishing them from conventional networks. Ref. [1] refers to these as ***flexible distribution networks*** (FDNs). Power electronic technology comprises FDNs with fascinating new features, and two of them are particularly concerned with operational flexibility enhancement: (*i*) it is technically feasible for distributed generators (DGs) to generate continuous active power and for static var compensators (SVCs) to generate continuous reactive power [2]; (*ii*) flexible networking devices, for example, soft open points (SOPs), have become a viable option to proactively adjust network power flows [3]. Consequently, FDNs can be considered competent players in providing both active and reactive power support to the upper-level transmission networks (TNs), which have been recognized to be in considerable need of such active and reactive power resources [4], [5].

The provision of active and/or reactive power from FDNs to TNs, which occurs at the transmission-distribution-interface (TDI), can be considered a type of flexibility resource for transmission system operators (TSOs). As demonstrated in [6], this TDI flexibility must be reliably assessed. Here, "***reliable***" refers to the range of the adjustable active and/or reactive power at the TDI, which is assessed by the distribution system operator (DSO) and then reported to the TSO upstream, must be deliverable whenever needed at the desired time scale. On the one hand, such a reliable TDI flexibility assessment helps the TSO with an enhanced capability of safely dispatching the TN. On the other hand, evaluating the TDI flexibility reliably ensures that the FDN can maintain its operational safety when adjusting the TDI power as required by the TSO upstream.

Refs. [6]-[17] studied the problem of evaluating TDI flexibility for vanilla distribution networks, where flexible networking devices are not fully considered. The main contributions of these works encompass (*i*) how to address the uncertainty in the power of DG and/or load [6]-[10], (*ii*) how to resolve the nonlinearity in the power flow and other operational constraints [10]-[12], and (*iii*) how to tactically model or relax the discrete optimal variables related to the tap position, switching of capacitor banks and operating of storage devices [10], [12], [13]-[16].

*In the first aspect*, random sampling methods can be adopted [7], [8], which can become computationally expensive and time-consuming as the number of samples increases. In contrast, in [6], the uncertainty is modeled as chance constraints, while [9], [10] resort to the robust optimization (RO) for a reliable assessment of the TDI flexibility under the uncertainty of DGs, and the resultant RO model can also be solved by iterative algorithms like column-and-constraint generation [10].

*In the second aspect*, [10]–[12] argue that a nearly three-phase-balanced condition holds in practice for high-voltage distribution networks; thus the classical Distflow equation and its linearized versions can be applied to model the network power flow and safety constraints as second-order conics [12] or linear constraints.

*In the third aspect*, [15] assumes a constant ratio, [16] relaxes the online load tap changer (OLTC) as a continuous controllable device to make the model easier to solve, and [10], [12], [13], [14] model the ratio of OLTCs and capacitor banks using integer variables or McCormick constraints, making the model more realistic.

*Lastly*, [6], [17] showcase the benefits that the TSO can leverage by using the TDI flexibility while preserving the safety of distribution networks. Nevertheless, the lack of

This work was supported by National Key R&D Program of China under grant 2022YFB2402900. S. Yang, Z. Li and Y. Tian are with the School of Electrical Engineering, Shandong University, Jinan 250061, China. Zhengshuo Li is the corresponding author (e-mail: zsli@sdu.edu.cn).

consideration of flexible networking devices makes the above studies unsuitable for FDNs, primarily because the enhanced operational flexibility of these devices cannot be captured and involved in the TDI flexibility assessment. Hence, extensive studies must be conducted.

Additionally, and what is more important, most studies adopt ***discrete-time (DT)*** models to assess TDI flexibility. Here, the DT model refers to an assessment model with the time granularity set to 5 min, 15 min, or 1 h, in alignment with the scheduling cycle of the power systems. The DT model assumes constant nodal power injection for each granularity period. Although this is a common assumption historically, recent works [18]–[20] have disclosed that owing to the prominent randomness and volatility of renewable generation, the related nodal power injection will experience considerable ***real-time*** fluctuations within one scheduling cycle, which cannot be accurately captured by existing TN dispatch models that are DT with a granularity of 5 min or longer. Hence, the scheduling decisions of the TSO are affected, leaving inappropriate reserves for subsequent adjustments.

This defect in the DT TN dispatch model has aroused interest in designing a ***continuous-time (CT)*** TN dispatch model for future renewable-dominated power systems [19], [20]. Regardless of whether this could replace the conventional DT TN dispatch model in the future, it is clear that to make a reasonable dispatch decision, the TSO should be aware of both the real-time variation in nodal power injection and the real-time systemwide flexibility, which includes TDI flexibility as an increasingly important component. Therefore, extant DT-TDI flexibility assessment models [6]–[17] may be inappropriate, and a CT alternative may become indispensable for this purpose. Moreover, a CT-TDI flexibility assessment model is conducive to safeguarding FDN operation, particularly when there exists a high penetration of DGs, primarily because the mode can reflect real-time safety while the FDN provides support to the upper-level TN.

To the best of the authors' knowledge, the majority of CT studies in the field of power system operation are concerned with economic dispatch and unit commitment problems, mostly on TNs [18]–[21]. None, or at least quite a few, of the studies have worked on a reliable CT assessment of TDI flexibility for FDNs. Moreover, as TDI flexibility is concerned with both active and reactive power, this assessment problem has to be modeled in a bi-objective pattern, which differs from most current CT studies that are only concerned with a single objective (perhaps after weighting). Notably, the regular weighted-sum approach with a set of given weights to reduce the bi-objective to a single objective, cannot be directly adopted for the concerned problem because the DSO should explore every possible pair of adjustable active and reactive powers at the TDI, which implies an infinite number of weights to be tested. In summary, the CT-TDI flexibility assessment problem is complicated and requires a new, tailored solution method.

To fill the identified gaps, this paper proposes a new assessment model of CT-TDI flexibility for FDNs and an efficient assessment algorithm that reliably assesses CT-TDI flexibility. The main contributions of this study are three-fold.

1) A novel notion of CT-TDI flexibility assessment for FDNs is introduced and a mathematical model is established. The differences between the proposed model and those in existing literature are summarized in [22]. In addition to accurately capturing the real-time situation of the FDN operation, this model can yield a reliable yet maximum TDI flexibility assessment by fully coordinating multisource flexibility from the DGs, energy storage systems (ESSs), and SVCs and harnessing flexible networking devices such as SOPs. The safety of FDN when delivering flexibility is also ensured by employing chance constraints to cope with uncertainty. Hence, the model can explore real-time flexibility at the TDI more effectively than regular DT assessment models.

2) To overcome the difficulty in constructing a quantitative metric that maximizes the range of both P and Q at any $t$ and the notion of CT, a new and proper metric is defined accordingly. This metric can tactically maximize the CT PQ output in the selected period such that the CT-TDI flexibility can be reasonably quantified.

3) A tailored and efficient solution method is designed to effectively solve the established model. It first recasts the bi-objective problem as a single-objective problem with a direction-factor-based approach, and then transforms the resultant infinite-dimension problem into a finite dimension problem that can be readily solved by off-the-shelf solvers. Moreover, this method can be implemented in parallel, significantly accelerating the solution process. Additionally, we provide an algorithm for visualizing decoupled PQ flexibility, which makes it easier for TSOs to oversee and leverage the single CT active and/or reactive TDI flexibility.

The remain of this paper is organized as follows. Section II proposes a mathematical model for CT-TDI flexibility assessment. Methods for obtaining CT-TDI flexibility and decoupled PQ flexibility are introduced in Section III. Case studies are presented in Section IV to verify the effects of the proposed method, and conclusions are presented in Section V.

## II. MATHEMATICAL MODEL OF THE CT-TDI FLEXIBILITY ASSESSMENT PROBLEM FOR FDN

### A. Problem Statement

Fig. 1 shows an example of an FDN and its connection with a TN through TDI. We assume that the FDN is equipped with multiple devices that can enhance its flexibility, such as DGs, ESSs, SOPs, SVCs, OLTCs, and voltage regulators, owing to modern power electronic technology.

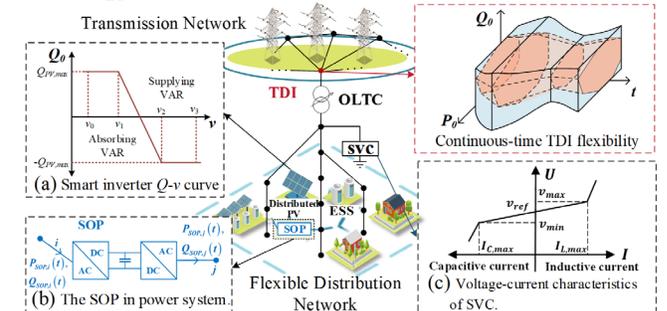

**Fig. 1.** An FDN connected with TN through TDI.

In this context, the CT-TDI flexibility assessment is used to quantify the maximum range of adjustable active and reactive powers at the TDI in terms of the CT resolution, while adhering to the CT FDN network safety alongside all the operating constraints of the devices against uncertain nodal power injections, where the devices include the smart inverters of DGs, ESSs, SOPs, SVCs, OLTCs, and voltage regulators.

**Remark 1**: Several studies [10], [23] suggest that high-voltage distribution networks that connect with the TN are approximately three-phase balanced in practice so that the classical Distflow [24] equation or its variations can be applied to model network safety constraints for simplicity. This convention is also followed in this work.

A system of notations is used in the subsequent modeling. Suppose there are N+1 nodes and N branches in an FDN. Let $\mathbb{N} = \{0, 1, \dots, N\}$ be the set of nodes, where Node 0 represents the *TDI* and $\mathbb{N}_{PV}, \mathbb{N}_{Load}$ is the set of nodes connected to the distributed photovoltaics (PV) units and load, respectively, and $\mathbb{L} = \{1, \dots, N\}$ be the set of branches.

To ensure that the main context remained concise and focused, only the essential constraints or the compact form of the model are presented. For a detailed examination, please refer to the complete form available in [22].

### B. Components of CT Operational Constraints

*1) Modeling the Distributed PV:* We assume that the distributed PV units considered in this study are equipped with a smart inverter interfaced with a grid. According to IEEE standards [25], this smart inverter enables the real-time reactive power output $Q_{PV,i}(t)$ of this PV unit, denoted by $i \in \mathbb{N}_{PV}$, to be regulated as a function of its connect node voltage $v_{PV,i}(t)$ as follows (also see the subplot Fig.1(a)),

$$Q_{PV,i}(t) = \begin{cases} Q_{PV}^{max}, & v_{PV,i}(t) \leq V_1 \\ \frac{2Q_{PV,max}}{V_1 - V_2}(v_{PV,i}(t) - V_1) + Q_{PV,max}, & V_1 \leq v_{PV,i}(t) \leq V_2, \\ -Q_{PV}^{max}, & v_{PV,i}(t) \geq V_2 \end{cases} \quad (1)$$

This nonlinear function (1) can be linearized by introducing type-2 special ordered set (SOS-2) variables $\{\lambda_{PV,i,j}(t)\}, j = 1,2,3,4$ as follows [26],

$$\begin{cases} Q_{PV,i}(t) = Q_{PV}^{max}(\lambda_{PV,i,1}(t) + \lambda_{PV,i,2}(t) - \lambda_{PV,i,3}(t) - \lambda_{PV,i,4}(t)), \\ U_{PV,i}(t) = \sum_{j=1}^{4} U_{i,j}\lambda_{PV,i,j}(t), \\ \forall t \in \mathcal{T}_{set}, \{\lambda_{PV,i,j}(t)\} \in SOS-2, \sum_{i=1}^{4} \lambda_{PV,i,j}(t) = 1, \lambda_{PV,i,j}(t) \geq 0. \end{cases} \quad (2)$$

In addition, the real-time active power $P_{PV,i}(t)$ of the PV unit is constrained by its maximum capacity $S_{PV}^{max}$, maximum forecast value $P_{PV,i}^{max}$ and the value of $Q_{PV,i}(t)$. Hence, the formulation of the feasible real-time operating set $\mathcal{X}_{PV}(t)$ related to the pair $(P_{PV,i}(t), Q_{PV,i}(t))$ of the $i$-th PV unit can be written as:

$$\mathcal{X}_{PV}(t) = \left\{ (P_{PV,i}(t), Q_{PV,i}(t)) \middle| \begin{array}{c} \exists (Q_{PV,i}(t), U_i(t)) \; satisfy \; (2) \\ [P_{PV,i}^2(t) + Q_{PV,i}^2(t) \leq (S_{PV}^{max})^2]_L \\ 0 \leq P_{PV,i}(t) \leq P_{PV,i}^{max} \end{array} \right\}, (3)$$

where $[\cdot]_L$ denotes the linearization of the expression $[\cdot]$ using the method in [22].

*2) Modeling the Load:* We consider that the active power $P_{Load,i}(t)$ of loads in $\forall i \in \mathbb{N}_{Load}$ are given, and assume a fixed power factor $\varphi_{Load,i}$ for $Q_{Load,i}(t)$ and $P_{Load,i}(t)$:

$$Q_{Load,i}(t) = \varphi_{Load,i} \cdot P_{Load,i}(t). \quad (4)$$

*3) Modeling the SOP:* Ref. [3] has demonstrated that owing to the use of back-to-back voltage source inverters, the SOP has the ability to flexibly regulate active and reactive power and isolate disturbances, as shown in Fig. 1(b). According to [3], constraints of the SOP include power balance constraints, reactive power constraints, and capacity limits. For $\forall (i,j) \in \mathbb{N}_{SOP}^2$, they can be modeled as follows:

$$\mathcal{X}_{SOP}^2(t) = \left\{ (\boldsymbol{P}_{SOP}(t)^T, \boldsymbol{Q}_{SOP}(t)^T) \middle| \begin{array}{c} \boldsymbol{A}_{SOP} \begin{bmatrix} \boldsymbol{P}_{SOP}(t) \\ \boldsymbol{Q}_{SOP}(t) \end{bmatrix} \leq \boldsymbol{c}_{SOP} \\ [P_{SOP,i}^2(t) + Q_{SOP,i}^2(t) \leq (S_{SOP}^{max})^2]_L \\ [P_{SOP,j}^2(t) + Q_{SOP,j}^2(t) \leq (S_{SOP}^{max})^2]_L \\ \boldsymbol{P}_{SOP}^{min} \leq \boldsymbol{P}_{SOP}(t) \leq \boldsymbol{P}_{SOP}^{max} \end{array} \right\}. \quad (5)$$

*4) Modeling the ESS:* Most ESS constraints include capacity, power, and complementarity constraints. For $\forall i \in \mathbb{N}_{ESS}$, they can be formulated as follows [27]:

$$\mathcal{X}_{ESS} = \left\{ (C_i(t), D_i(t)) \middle| \begin{array}{cc} 0 \leq E_s(t_0) - & \\ \int_{t_0}^{t} \left( D_i(t)\frac{P_{D,i}}{\eta_D} - C_i(t)\eta_C P_{C,i} \right) dt \leq E_s^{max} & (6a) \\ 0 \leq C_i(t) \leq 1, 0 \leq D_i(t) \leq 1 & (6b) \\ C_i(t)D_i(t) = 0 & (6c) \end{array} \right\}.$$

However, it should be noted that unlike the DT counterpart, simply imposing the CT constraint (6) may cause an instantaneous and overly frequent transition of the charge-discharge behavior, which is undesirable and unrealistic. Hence, we have to impose the following constraint (7), which restricts minimum charge/discharge time to prevent that effect, where $t_k$ represents the time when values of $C_i(t)$ and $D_i(t)$ change, and $sgn(\cdot)$ represents the sign function:

$$\int_{t_k}^{t_k+T_{ESS}^C} sgn(C_i(t))dt \geq T_{ESS}^C, \int_{t_k}^{t_k+T_{ESS}^D} sgn(D_i(t))dt \geq T_{ESS}^D. \quad (7)$$

Complementarity constraints (6c) and (7) are non-convex constraints, making the model difficult to solve. To relax these, we assume $C_i(t) = 1 - D_i(t)$ [28], and the model becomes:

$$\mathcal{X}_{ESS} = \left\{ D_i(t) \middle| \begin{array}{cc} \int_{t_0}^{t} (\boldsymbol{A}_{ESS}D_i(t) + \boldsymbol{c}_{ESS1})dt \leq \boldsymbol{c}_{ESS2} & (8a) \\ 0 \leq D_i(t) \leq 1 & (8b) \\ \int_{t_k}^{t_k+T_{ESS}^C} sgn(1 - D_i(t))dt \geq T_{ESS}^C & \\ \int_{t_k}^{t_k+T_{ESS}^D} sgn(D_i(t))dt \geq T_{ESS}^D & (8c) \end{array} \right\},$$

where $\boldsymbol{A}_{ESS}, \boldsymbol{c}_{ESS1}$, and $\boldsymbol{c}_{ESS2}$ are the coefficient matrixes and vectors in the compact form, respectively. Constraint (8c) is addressed in Section III by assuming an invariant charging/discharging status during the selected period.

*5) Modeling the SVC:* Ref. [29] shows that as the system voltage decreases, the ability of the SVC to output a reactive current also decreases. Fig. 1(c) shows the voltage-current characteristics of the SVC. Define $U = v^2$, for $\forall i \in \mathbb{N}_{SVC}$, it is modeled as follows:

$$\mathcal{X}_{SVC}(t) = \left\{ Q_{SVC,i}(t) \middle| Q_{SVC,i}(t) = \frac{1}{2}k_{SVC}(U_i(t) - U_{ref}) \right\}. \quad (9)$$

where $k_{SVC}$ denotes the rate of change of the SVC current with voltage and $U_{ref}$ denotes the square value of the reference voltage of the SVC.

*6) Modeling the Capacitor Shunt:* According to [10], $\forall i \in \mathbb{N}_C$, the output $Q_{C,i}(t)$ of a capacitor shunt is modeled and linearized using McCormick envelope relaxation as follows:

$$\mathcal{X}_C(t) == \left\{ Q_{C,i}(t) \left| \begin{array}{l} Q_{C,i}(t) = \sum_k q_{Ck} z_{c,i,k}(t) \\ U_i(t) - \left(1 - z_{c,i,k}(t)\right) U_i^{max} \leq z_{c,i,k}(t) \\ U_i(t) - \left(1 - z_{c,i,k}(t)\right) U_i^{min} \geq z_{c,i,k}(t) \\ \lambda_{c,i,k}(t) U_i^{min} \leq z_{c,i,k}(t) \leq \lambda_{c,i,k}(t) U_i^{max} \\ \forall t \in \mathcal{T}_{set}, \{\lambda_{c,i,k}(t)\} \in SOS-1 \\ \sum_k \lambda_{c,i,k}(t) = 1, \lambda_{c,i,k}(t) \geq 0 \end{array} \right. \right\}, \quad (10)$$

where $q_{Ck}$ denotes its capacitance, and $z_{c,i,k}(t)$ and $\lambda_{c,i,k}(t)$ are the variables introduced because of the relaxation and selecting $q_{Ck}$.

*7) Modeling the FDN Network Constraints:* Following Remark 1, a linearized Distflow model is used to construct the load flow. To facilitate modeling, branches $\mathbb{L}$ are divided into three sets: $\mathbb{L}_{OLTC}, \mathbb{L}_{Reg}$ and $\mathbb{L}_{rest}$. Here, $\mathbb{L}_{OLTC}$ is the set of branches with an OLTC, $\mathbb{L}_{Reg}$ is the set with a voltage regulator, and $\mathbb{L}_{rest}$ is the set with the remaining branches.

$$\begin{cases} P_{L,i}(t) = P_{Load,i}(t) - P_{PV,i}(t) + D_i(t) \cdot P_{D,i} - (1 - D_i(t)) \cdot P_{C,i} \\ \quad + P_{SOP,i}(t) = \sum_{k \in \mathcal{C}(i)} P_{ik} - \sum_{k \in \mathcal{P}(i)} P_{ik}, \\ Q_{L,i}(t) = Q_{Load,i}(t) - Q_{PV,i}(t) + Q_{SOP,i}(t) - Q_{SVC,i}(t) \\ \quad - Q_{C,i}(t) = \sum_{k \in \mathcal{C}(i)} Q_{ik} - \sum_{k \in \mathcal{P}(i)} Q_{ik}, \\ \forall (i,j) \in \mathbb{L}_{OLTC}: \\ U_i(t) - U_{OLTC,j}(t) = 2\left(r_{ij} \cdot P_{ij}(t) + x_{ij} \cdot Q_{ij}(t)\right), \\ U_{OLTC,j}(t) = \sum_k a_{OLTC,k}^2 \lambda_{OLTC,i,k}(t), \\ U_{OLTC,j}(t) - \left(1 - z_{OLTC,j,k}(t)\right) U^{max} \leq z_{OLTC,j,k}(t), \\ U_{OLTC,j}(t) - \left(1 - z_{OLTC,j,k}(t)\right) U^{min} \geq z_{OLTC,j,k}(t), \\ \lambda_{OLTC,j,k}(t) U^{min} \leq z_{OLTC,j,k}(t) \leq \lambda_{OLTC,j,k}(t) U^{max}, \\ \forall t \in \mathcal{T}_{set}, \{\lambda_{OLTC,j,k}(t)\} \in SOS-1, \\ \sum_k \lambda_{OLTC,j,k}(t) = 1, \lambda_{OLTC,j,k}(t) \geq 0, \\ \begin{cases} \forall (i,j) \in \mathbb{L}_{Reg}: \\ U_i(t) - U_j(t) = 2\left(r_{ij} \cdot P_{ij}(t) + x_{ij} \cdot Q_{ij}(t)\right), \\ (\tau^{min})^2 U_j(t) \leq U_i(t) \leq (\tau^{max})^2 U_j(t), \end{cases} \\ \begin{cases} \forall (i,j) \in \mathbb{L}_{rest}: \\ U_i(t) - U_j(t) = 2\left(r_{ij} \cdot P_{ij}(t) + x_{ij} \cdot Q_{ij}(t)\right), \\ U^{min} \leq U_i(t) \leq U^{max}. \end{cases} \end{cases} \quad (11)$$

$P_{L,i}(t)/Q_{L,i}(t)$ represent active/reactive net power injection at node $i$; $P_{ij}(t)$ and $Q_{ij}(t)$ represent the active and reactive power flowing through branches; $\mathcal{P}(i)$ and $\mathcal{C}(i)$ are the sets of the parent and child nodes of node $i$; $a_{OLTC,k}^2$ is the square of the possible ratio of OLTC; $\tau_{min}$ and $\tau_{max}$ represent the minimum and maximum tap position of the voltage regulator respectively. For branches with an OLTC, similar to (10), auxiliary variables $\lambda_{OLTC,j,k}(t)$ and $z_{OLTC,i,k}(t)$ are used to deal with the bilinear term.

### C. The Model of CT-TDI Flexibility under Uncertainty

This study considers the uncertainty in PV units and loads and follows the assumption in [30] that the uncertainty follows a Gaussian distribution and is independently and identically distributed with each other:

$\forall i \in \mathbb{N}_{PV}$,
$$P_{PV,i}^{max}(t) = P_{PV0,i}^{max}(t) + \tilde{P}_{PV,i}, \tilde{P}_{PV,i} \sim N(0, \sigma_{PV,i}^2).$$

$\forall i \in \mathbb{N}_{Load}$,
$$P_{Load,i}(t) = P_{Load0,i}(t) + \tilde{P}_{Load,i}, \tilde{P}_{Load,i} \sim N(0, \sigma_{Load,i}^2).$$

The uncertainty set can be summarized as:

$$\mathcal{U} = \left\{ [[\tilde{P}_{PV,i}]; [\tilde{P}_{Load,i}]] \left| \begin{array}{l} \tilde{P}_{PV,i} \sim N(0, \sigma_{PV,i}^2), \forall i \in \mathbb{N}_{PV} \\ \tilde{P}_{Load,i} \sim N(0, \sigma_{Load,i}^2), \forall i \in \mathbb{N}_{Load} \end{array} \right. \right\}.$$

Considering the model predefined, the variables are divided into three types: Uncertain variables $\boldsymbol{u} = [[\tilde{P}_{PV,i}]; [\tilde{P}_{Load,i}]]$, independent decision variables $\boldsymbol{x}(t) = [[P_{PV,i}(t)]; [D_i(t)]; [\lambda_{c,i,k}(t)]; [\boldsymbol{P}_{SOP}(t)^T, \boldsymbol{Q}_{SOP}(t)^T]^T; [\lambda_{OLTC,i,k}(t)]]$, and dependent variables $\boldsymbol{y}(t) = [[Q_{PV,i}(t)]; [Q_{SVC,i}(t)]; [U_i(t)]; [P_{ij}(t)]; [Q_{ij}(t)]; [z_{c,i,k}(t)]; [z_{OLTC,i,k}(t)]]$ that values of which are determined when $\boldsymbol{u}$ and $\boldsymbol{x}(t)$ are decided.

Suppose a suitable metric $M(T)$ is defined to reasonably quantify the flexibility for any $t \in T = [t_1, t_2]$, which will be investigated in Section II D. Subsequently, the compact form of overall CT-TDI flexibility under uncertainty can be represented as follows:

$$\max M(T) \quad (12a)$$
$$P_0(t) = \boldsymbol{A}_{P0}\boldsymbol{y}(t), Q_0(t) = \boldsymbol{A}_{Q0}\boldsymbol{y}(t),$$
$$\forall t \in T, \boldsymbol{u} \in \mathcal{U}, \exists \boldsymbol{y}(t): \begin{cases} \boldsymbol{A}\boldsymbol{x}(t) + \boldsymbol{B}\boldsymbol{y}(t) + \boldsymbol{F}\boldsymbol{u} = \boldsymbol{c}, \\ \boldsymbol{C}\boldsymbol{x}(t) + \boldsymbol{D}\boldsymbol{y}(t) + \boldsymbol{G}\boldsymbol{u} \leq \boldsymbol{d}, \end{cases} \quad (12b)$$

Bold uppercase letters denote coefficient matrixes whereas the bold letters denote vectors in the constraints. The detailed information regarding the constraints could be found in [22].

Reformulate the uncertain constraints in (12b) as chance constraints below:

$$\forall t \in T, \boldsymbol{u} \in \mathcal{U}, \exists \boldsymbol{y}(t): \begin{cases} \boldsymbol{A}\boldsymbol{x}(t) + \boldsymbol{B}\boldsymbol{y}(t) = \boldsymbol{c}, \\ \mathbb{P}\mathrm{r}[\boldsymbol{C}\boldsymbol{x}(t) + \boldsymbol{D}\boldsymbol{y}(t) \leq \boldsymbol{d}]_{\mathbb{R}} \geq 1 - \alpha, \end{cases} \quad (13)$$

$\mathbb{P}\mathrm{r}[\cdot]_{\mathbb{R}}$ represents the probability of the constraints in $[\cdot]$. The confidence-level $\alpha$ is the maximum allowable probability of constraints violation.

It follows [30] that the above chance constraints can be safely transformed into deterministic constraints, such that the original uncertain model (12) can be recast as follows:

$$\max M(T) \quad (14a)$$
$$\forall t \in T: \begin{cases} P_0(t) = \boldsymbol{A}_{P0}\boldsymbol{y}(t), Q_0(t) = \boldsymbol{A}_{Q0}\boldsymbol{y}(t), \\ \boldsymbol{A}\boldsymbol{x}(t) + \boldsymbol{B}\boldsymbol{y}(t) = \boldsymbol{c}, \\ \boldsymbol{C}\boldsymbol{x}(t) + \boldsymbol{D}\boldsymbol{y}(t) \leq \bar{\boldsymbol{d}}, \end{cases} \quad (14b)$$

The detailed information regarding the reformulation could be found in [22].

### D. Establishing Objective Function M(t)

The objective function is a metric for quantifying the maximum adjustable range of the active and reactive powers at the TDI:

$$\max_{\boldsymbol{x}(t)} M(T) = Range(P_0(t), Q_0(t)) \quad for \ t \in T. \quad (15)$$

The decision variable is $\boldsymbol{x}(t)$. However, as shown in Fig. 2, there are two challenges in constructing the metric in (15).

The first is, with a fixed $t$, how to optimize the maximum adjustable range of $P_0(t)$ and $Q_0(t)$, i.e., how to optimize the area that satisfies the constraints in (10). In particular, since the FDN naturally does not satisfy $X \gg R$ and grid operational constraints (14b), $P_0(t)$ and $Q_0(t)$ are coupled so the area is generally a polygon instead of a rectangular, particularly under the impact of uncertain PV output and load fluctuations. Therefore, the selected metric must reflect this coupling reasonably.

Second, the notion of the CT in Fig. 2(a) mathematically makes (14) an infinite-dimensional problem. With the addition of the time axis, the two-dimensional P-Q region extends and transforms into a three-dimensional tube. Hence, the selected metric must reasonably reflect the shape of a tube formed by countless polygons related to $t$ within a continuous time interval $[t_1, t_2]$, and maximize the tube.

Inspired by [13], which uses a direction-factor-based function for the assessment of active and reactive power flexibility in the DT-TDI, a solution to the aforementioned difficulties is designed as follows. The key idea is to translate (15) into problem (16) by introducing a varying parameter $\theta$:

$$\text{for all } \theta \in [0, 2\pi], \text{ solve:}$$
$$Obj := \max_{x(t)} \int_{t_1}^{t_2} S_0(t) dt,$$
$$s.t. \forall t \in [t_1, t_2] \begin{cases} P_0(t) = S_0(t)\cos\theta, Q_0(t) = S_0(t)\sin\theta, \\ (10b). \end{cases} \quad (16)$$

Equation (16) is illustrated in Fig. 2(c). Starting from the origin of the PQ cross-section at time $t_1$, for any given $\theta$, draw the cross-section related to $\theta$ intersected with the tube until $t_2$, (16) aims to maximize the area of this cross-section, which can be written as $A(\theta)$. Moreover, after solving $A(\theta)$, a *slice-$\theta_k$* can be obtained, whose curve boundary reflects the optimal $S_0(t)$ that satisfies (12) under the given $\theta$. If $\theta$ is discretized, we obtain discrete $A(\theta)$ values. The regions between two adjacent $A(\theta)$ are filled by the affine combination of both adjacent $A(\theta)$ with respect to $\theta$. Ultimately, we obtain the optimal $S_0(t)$ in any direction within time $T$, which represents TDI flexibility from $t_1$ to $t_2$.

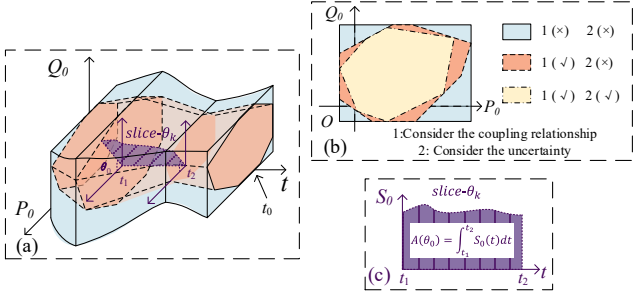

**Fig. 2.** (a) CT-TDI flexibility. (b) Influence of the coupling relationship and the uncertainty for the TDI flexibility. (c) Meaning of (12).

## III. SOLUTION METHODS FOR OBTAINING CT-TDI FLEXIBILITY AND DECOUPLED PQ FLEXIBILITY

### A. The Overall Solving Framework

Equation (12) is difficult to optimize because (*i*) (16) evaluates all $\theta \in [0, 2\pi]$ so that DSO could get every possible pair of adjustable active and reactive power at the TDI, which implies an infinite number of $\theta$. (*ii*) For a fixed $\theta$, the variables and the constraints are infinite dimensions, making $A(\theta)$ difficult to solve.

The following solution framework is designed in response to the two questions: First, the continuous $\theta \in [0, 2\pi]$ are discretized into a limited number of $\theta_k$ such that the subproblems evaluated at these $\theta_k$ can be solved in parallel.

Second, for every subproblem $A(\theta_k)$ related to a given $\theta_k$, the BP spline decomposition [19] is utilized to transform the infinite-dimensional CT problem into a finite-dimensional mixed-integer linear programming problem that can be solved using existing commercial solvers.

Finally, after solving $A(\theta_k)$, a *slice-$\theta_k$* can be obtained. As shown in Fig. 3, all *slice-$\theta_k$* and their linear affine combinations form the tube of $P_0(t)$ and $Q_0(t)$ within the given interval $[t_1, t_2]$ in any direction with respect to $t$. The rule related to the linear affine combination is introduced detail in Section III C (15). The corresponding $P_0(t_2)$ and $Q_0(t_2)$ values can also be obtained at a specific time $t_2$. We considered the tube to exhibit the CT-TDI flexibility.

Fig. 4 shows a flowchart of the solution method.

In the next section, we introduce the transformation of an infinite-dimensional problem into a finite-dimensional solvable problem after $\theta_k$ is given. Subsequently, the discretization of $\theta$ used in this paper and the procedures involved in the third stage are briefly discussed.

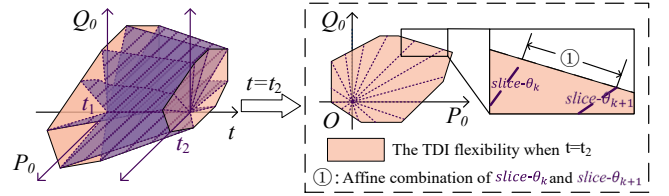

**Fig. 3.** Method to obtain the overall TDI flexibility.

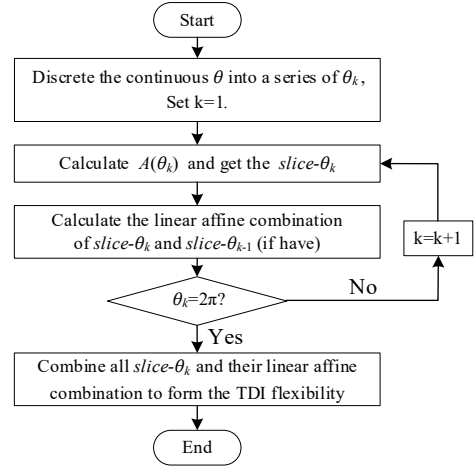

**Fig. 4.** Flowchart of the solution method.

### B. Transformation of the Subproblem $A(\theta_k)$ Using BP Spline Decomposition

In this section, BP spline decomposition is used to transform an infinite-dimensional CT model into a finite-dimensional model. The basic BP spline mathematical properties including the integral term transformation, inequality term transformation and equality term transformation, ensure the equivalence of the transformation. Further details on the BP method and its properties can be found in [19].

Divide the time period $[t_1, t_2]$ into m segments with the same scheduling cycle $T$, i.e., $mT = t_2 - t_1$, and the starting time of each interval is $t_m = t_1 + (m-1)T$. To make the problem easier to solve, we assume that during one scheduling cycle, the charging or discharging state of the ESS remains unchangeable, making (8c) naturally satisfied. The subscripts B0, B1, B2, and B3 represent the four control coefficients of a physical quantity within a period, whereas the subscripts m=1, 2, ..., denote the period in which these control coefficients are found. The objective function of $A(\theta_k)$ can be written as

$$Obj := \max_{\theta_k} \int_{t_1}^{t_2} S_0(t) dt = \max \int_{t_1}^{t_2} \sum_m \boldsymbol{S}_{\boldsymbol{0,B},\boldsymbol{m}}^T \boldsymbol{B}_{\boldsymbol{3,m}}(t) dt, \quad (17)$$

where
$$\boldsymbol{S_{0,B,m}^T} = [S_{0,B0,m}, S_{0,B1,m}, S_{0,B2,m}, S_{0,B3,m}]^T,$$

$$\boldsymbol{B_{3,m}}(t) = \boldsymbol{B_{3,m}}\left(\frac{t-t_m}{T}\right) = \begin{bmatrix} C_3^0\left(1-\frac{t-t_m}{T}\right)^3 \\ C_3^1\left(\frac{t-t_m}{T}\right)\left(1-\frac{t-t_m}{T}\right)^2 \\ C_3^2\left(\frac{t-t_m}{T}\right)^2\left(1-\frac{t-t_m}{T}\right) \\ C_3^3\left(\frac{t-t_m}{T}\right)^3 \end{bmatrix}.$$

The constraints in (16) can also be transformed using the BP spline properties.

Furthermore, objective function (17) can be divided into $m$ sub-objective functions according to $T$:

$$Obj := \max_{\theta_k,m} \max_{\boldsymbol{x_{B,m}}} \int_{t_m}^{t_m+T} \boldsymbol{S_{0,B,m}^T} \boldsymbol{B_{3,m}}(t)\,dt$$
$$= \max_{\boldsymbol{x_{B,m}}} \boldsymbol{S_{0,B,m}^T} \int_{t_m}^{t_m+T} \boldsymbol{B_{3,m}}(t)\,dt = \max_{\boldsymbol{x_{B,m}}} \frac{T \cdot \boldsymbol{1}^T \boldsymbol{H_{0,B,m}}}{4}.$$

The transformed constraints can also be classified according to $T$. Combining the sub-objective function and constraints with the same $T$ respectively, the sub-question $A(\theta_k)$ is divided into $m$ optimization problems to accelerate the solving progress. Each optimization problem has the following form:

$$Obj := \max_{\theta_k,m} \max_{\boldsymbol{x_{B,m}}} \frac{T \cdot \boldsymbol{1}^T \boldsymbol{S_{0,B,m}}}{4},$$
$$s.t.$$
$$\boldsymbol{P_{0,Bl,m}} = \boldsymbol{A_{B,P0}} \boldsymbol{y_{B,m}}, \boldsymbol{Q_{0,Bl,m}} = \boldsymbol{A_{B,Q0}} \boldsymbol{y_{B,m}},$$
$$\boldsymbol{A_B} \boldsymbol{x_{B,m}} + \boldsymbol{B_B} \boldsymbol{y_{B,m}} = \boldsymbol{c_B}, \quad (18)$$
$$\boldsymbol{C_B} \boldsymbol{x_{B,m}} + \boldsymbol{D_B} \boldsymbol{y_{B,m}} \le \bar{\boldsymbol{d}}_B,$$
$$\boldsymbol{P_{0,B,m}} = \cos\theta_k \cdot \boldsymbol{S_{0,B,m}}, \boldsymbol{Q_{0,B,m}} = \sin\theta_k \cdot \boldsymbol{S_{0,B,m}}.$$

The optimization variables in the model include:
$$\boldsymbol{x_{B,m}} = [\boldsymbol{x_{B0,m}}; \boldsymbol{x_{B1,m}}; \boldsymbol{x_{B2,m}}; \boldsymbol{x_{B3,m}}].$$

And $\boldsymbol{A_{B,P0}}, \boldsymbol{A_{B,Q0}}, \boldsymbol{A_B}, \boldsymbol{B_B}, \boldsymbol{C_B}, \boldsymbol{D_B}$ are coefficient matrixes after transmission; $\boldsymbol{c_B}, \bar{\boldsymbol{d}}_B$ are coefficient vectors after transmission. More information can be found in [22].

Notably, with the fixed $\theta_k$, the obtained problem (14) is a mixed-integer linear programming problem, which is solvable with commercial solvers. The computational time test is presented in Section IV B.

Upon solving $A(\theta_k)$, we use the inverse transformation transforming all $\boldsymbol{S_{0,B,m}}$ into to $S_0(t)$ by (17) to obtain a $slice$-$\theta_k$.

### C. The Discretization of $\theta$ and Solving Framework

For the continuous $\theta$, we use a uniform sampling method to sample $\theta$, generalizing the set of $\theta_k$ whose value varies from 0 to $\pi$. For each $\theta_k$, the corresponding $slice$-$\theta_k$ and $slice$-$(\theta_k + \pi)$ are computed following the procedure outlined in Section III B. The flexibility of $P_0(t)$ and $Q_0(t)$ under $\theta_k$ can be calculated according to (19):

$$Up\ flexibility: \begin{cases} P_0^{\theta_k}(t) = S_0^{\theta_k}(t)\cos\theta_k, \\ Q_0^{\theta_k}(t) = S_0^{\theta_k}(t)\sin\theta_k, \end{cases}$$
$$Down\ flexibility: \begin{cases} P_0^{\theta_k+\pi}(t) = S_0^{\theta_k+\pi}(t)\cos(\theta_k+\pi), \\ Q_0^{\theta_k+\pi}(t) = S_0^{\theta_k+\pi}(t)\sin(\theta_k+\pi). \end{cases} \quad (19)$$

For $\forall \theta \in [\theta_{k-1}, \theta_k]$, choose $slice$-$\theta_k$ and $slice$-$\theta_{k+1}$ at the interval endpoints to generate the linear affine combination with any $\theta$:

$$P_0^\theta(t) = \frac{\theta_k - \theta}{\theta_k - \theta_{k-1}} P_0^{\theta_{k-1}}(t) + \frac{\theta - \theta_{k-1}}{\theta_k - \theta_{k-1}} P_0^{\theta_k}(t),$$
$$Q_0^\theta(t) = \frac{\theta_k - \theta}{\theta_k - \theta_{k-1}} Q_0^{\theta_{k-1}}(t) + \frac{\theta - \theta_{k-1}}{\theta_k - \theta_{k-1}} Q_0^{\theta_k}(t). \quad (20)$$

The flexibility under $\theta$ can also be calculated according to (19). Additionally, all $slice$-$\theta_k$ and their linear affine combination form a tube and we believe that the tube is the CT-TDI flexibility within the given interval. By substituting a specific $t_2$ into (19) and (20), we obtain the PQ flexibility at moment $t_2$, as shown in Fig. 5(a).

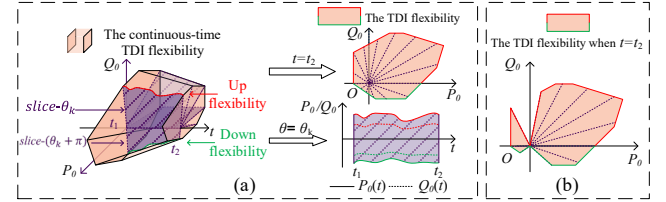

**Fig. 5.** (a) CT-TDI flexibility and obtain the flexibility at any $t$ and $\theta$. (b) TDI flexibility at $t_2$ when existing unsolvable $\theta$.

**Remark 2**: In practical situations, it is possible that under certain load inputs, the FDN cannot achieve any power factor output on the TDI by adjusting its own generation and flexibility devices. This could result in some $\theta$ corresponding to (18) being unsolvable, as shown in Fig. 5(b). Therefore, the range of TDI flexibility at a specific moment may not necessarily be convex and could potentially have gaps.

### D. An Algorithm for Visualizing Decoupled PQ Flexibility

We have already obtained a PQ flexibility diagram for a specific time $t$. In practical dispatch operations, we also notice that practical TSOs may also be concerned with the range of adjustable PQ from the FDNs [4] [5]. A chart that visualizes this decoupled PQ flexibility facilitates the TSO in making certain dispatch decisions. In this section, we introduce an effective algorithm for visualizing decoupled PQ flexibility.

The core concept of this algorithm is to select an appropriate initial point $(P_0, Q_0)$ with a specific $t_0$. In conjunction with Remark 2, considering that some $\theta$ values may be infeasible, we discuss the following three conditions with different shapes of TDI flexibility:

(*i*) For the TDI flexibility existing in all $\theta$, $P_0 = 0.5 \cdot \max_\theta(S_0^{\theta_k}(t_0) + S_0^{\theta_k+\pi}(t_0)) \cdot \cos\theta_k$, $Q_0 = 0.5 \cdot \max_\theta(S_0^{\theta_k}(t_0) + S_0^{\theta_k+\pi}(t_0)) \cdot \sin\theta_k$.

For the TDI flexibility existing in some $\theta$, choose the biggest piece by counting the number of $\theta_k$ in this piece, and:

(*ii*) If in the largest piece, $\max\theta_k - \min\theta_k \le \pi$, $P_0 = 0.5 \cdot S_0^{0.5 \cdot (\max\theta_k - \min\theta_k)}(t_0) \cdot \cos\theta_k, Q_0 = S_0^{0.5 \cdot (\max\theta_k - \min\theta_k)}(t_0) \cdot \sin\theta_k$.

(*iii*) If in the largest piece, $\max\theta_k - \min\theta_k > \pi$, $P_0 = 0.5 \cdot \max_\theta S_0^{\theta_k}(t_0) \cdot \cos\theta_k, Q_0 = 0.5 \cdot \max_\theta S_0^{\theta_k}(t_0) \cdot \sin\theta_k$. These three types are illustrated in Fig. 6.

Upon obtaining the initial points, the following algorithm can be initiated, which is adapted from [31]. It employs an exploratory approach to expand the feasible region incrementally. This expansion continues until further enlargement is not possible in any direction. Algorithm 1 provides the pseudocode for the proposed decoupled PQ flexibility visualization algorithm.

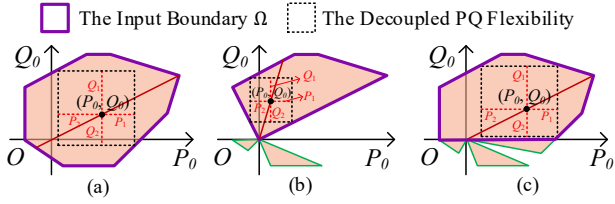

**Fig. 6.** (a) TDI flexibility existing in all $\theta$. (b) The largest piece of TDI flexibility satisfying $\max \theta_k - \min \theta_k \leq \pi$. (c) The largest piece of TDI flexibility satisfying $\max \theta_k - \min \theta_k > \pi$.

---

**Algorithm 1** The Decoupled PQ Flexibility Visualization Algorithm

**Input:** The initial point $(P_0, Q_0)$; The initial step sizes $\Delta$; Allowable error $\epsilon$; the boundary $\Omega$; Directional multiplier $k_1 = 1, k_2 = -1, k_3 = 1, k_4 = -1$; Initial point parameters $P_1 = P_0, P_2 = P_0, Q_1 = Q_0, Q_2 = Q_0$.

**Output:** Final size $(P_1, P_2, Q_1, Q_2)$

1  Calculate $P_1 = P_1 + k_1\Delta$, $P_2 = P_2 + k_2\Delta$, $Q_1 = Q_1 + k_3\Delta$, $Q_2 = Q_2 + k_4\Delta$
2  **if** $(P_1, Q_1) \in \Omega, (P_1, Q_2) \in \Omega, (P_2, Q_1) \in \Omega$ and $(P_2, Q_2) \in \Omega$
3      **then go to** 1
4  **elseif** $k_i \leq \epsilon, i = \{1,2,3,4\}$ for points that do not meet the requirements
5      **then** $P_i = P_i - k_i\Delta, Q_i = Q_i - k_i\Delta$, set $k_i = 0$, **go to** 8
6  **else**
        $k_i = k_i/10$ for points that do not meet the requirements, $P_i = P_i - k_i\Delta, Q_i = Q_i - k_i\Delta$, **go to** 1
7  **end**
8  **if** $k_1 = 0, k_2 = 0, k_3 = 0$ and $k_4 = 0$
9      **then** stop the iteration
10 **else**
        **go to** 1
11 **end**

---

## IV. CASE STUDIES

### A. Test Systems

In this section, the proposed mathematical model and solution methods are tested on a 12-node test system, 141-bus system, and a 533-bus system.

Based on the line parameters in [32], a 12-node system is implemented to analyze the feasibility and performance of the proposed model. As shown in Fig. 7, the 12-node system has an OLTC between node 1 and the upper-level TN. Three distributed PVs and ESSs were placed at nodes 5,9, and 12. Nodes 3, 5, 7, 8, 9, 10 and 12 are load nodes and $\sigma_{PV}^2 = 0.005$, $\sigma_{Load}^2 = 0.001$. Two SVCs were placed at nodes 2 and 6. The capacitor shunt was placed at node 6. An SOP was installed between nodes 3 and 6. A voltage regulator was installed between nodes 3 and 4. A detailed description of the system parameters is available in [22].

In addition, the modified IEEE 141-bus distribution system [32] and modified IEEE 533-bus distribution system [33] were used to test the stability of the model in large-scale cases. Five distributed PVs were placed at nodes 50, 52, 67, 68, 85, and 87, respectively, and an SOP was placed between nodes 6 and 69 in the 141-bus system. Similarly, a distributed PV was placed at node 254 and an SOP was placed between node 8 and 67 for the 533-bus system. Further information regarding the test systems is available in [22].

In this section, we test the CT-TDI flexibility for an hour and choose the scheduling cycle $T$ to be 15 min. To make the simulation more realistic, we considered the load from the modified Tetuan City power-consumption data [34], as shown in Fig. 8(a). The predicted CT maximum distributed PV output is illustrated in Fig. 8(b), which was obtained from [34]. Unless otherwise specified, the PV outputs and loads in the following simulations were initialized according to Fig. 8.

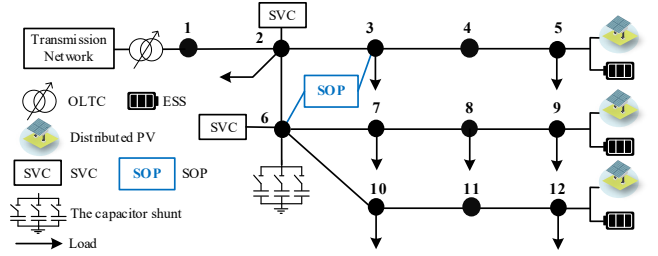

**Fig. 7.** Diagram of the 12-node test system.

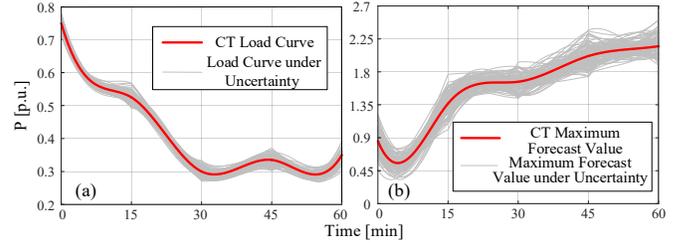

**Fig. 8.** (a) Predicted curve of the total load. (b) Predicted maximum output curve of the total distributed PVs

Based on the three test systems above, the following cases were tested: The first is designed to prove the correctness of the proposed model and solution method. Set $\alpha = 0.1$, the results from CT model are compared with those from DT model, demonstrating the superiority of the CT model. The DT model is established based on the CT model, and detailed information can be found in [22]. Additionally, the effect of the decoupled PQ flexibility visualization algorithm is tested.

In the second case, we want to show the influence of relevant factors such as SOP, PV penetration levels, use of ESS, and uncertainties on TDI flexibility. By testing TDI flexibility of the system with and without SOP, changing the PV penetration level and ESS capacity, or varying values of $\alpha$ and $\sigma_{PV}^2$, $\sigma_{Load}^2$ respectively, we observe the changes of TDI flexibility.

All case studies were conducted in MATLAB R2020a on an Intel (R) Core (TM) i7-10700F, 2.9 GHz, 16 GB RAM PC using CPLEX 12.10.0.0.

### B. Compare the CT-TDI Flexibility with DT-TDI Flexibility

We arbitrarily select the PQ flexibility of the 12-node and 141-bus systems at a specific time, as shown in Fig. 9(a) and Fig. 9(b). the net load curves throughout the test period are plotted in Fig. 10(a). Each figure presents a comparison of the evaluation results of the CT and DT methods. Moreover, we compare the CT-TDI flexibility and the DT-TDI flexibility under specific $\theta$, as depicted in Fig.10(b).

Several observations are made based on Fig. 9 and Fig.10: In Fig. 9(a) and Fig. 9(b), notable differences can be observed between the results obtained from the CT and DT evaluations. These variations suggest that, despite the increased complexity associated with the CT model, its use is essential. Second, a comparison of Fig. 10(a) and 10(b) reveals that the up flexibility defined in (19) is negatively correlated with the net load, whereas the down flexibility is positively correlated. Third, owing to network constraints, PQ flexibility remains coupled, regardless of whether the CT or DT methods are used.

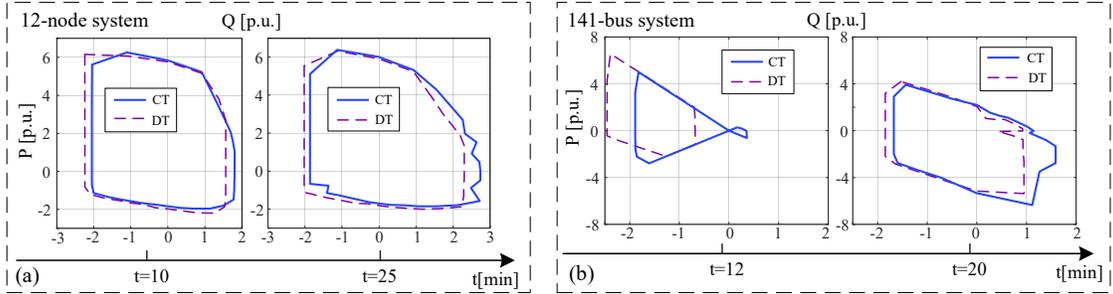

**Fig. 9.** (a) CT-, DT-TDI flexibility of the 12-node system when $t = 10, 25$. (b) CT-, DT-TDI flexibility of the 141-bus system when $t = 12, 20$.

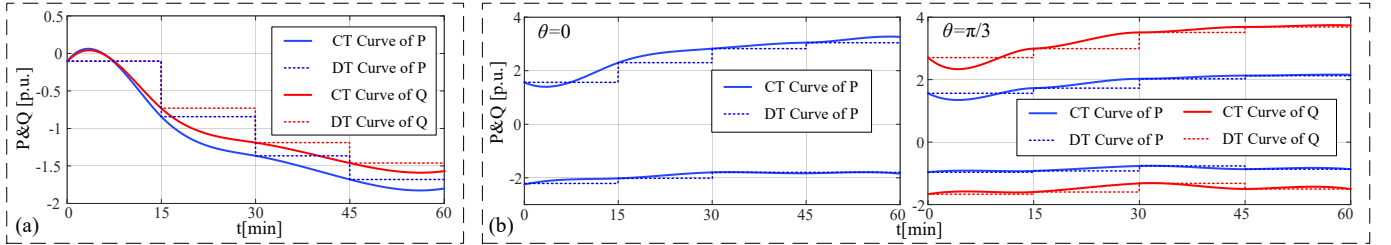

**Fig. 10.** (a) CT and DT net load curves. (b) CT- and DT-TDI flexibility when $\theta = 0, \frac{\pi}{3}$ for the 12-node system.

Furthermore, from Fig. 10, it can be observed that when $t = 12$, during the simulation, the system becomes unsolvable at some $\theta$ values, such as $\theta = \pi/2$. This indicates that the FDN cannot support the upper-level TN with such $\theta$, which substantiates Remark 2. Upon inspection, it is found that a lack of PV generation led to the problem being unsolvable. These cases underscore the need for TSOs to ensure that FDN flexibility is accurately assessed when dispatch decisions are issued. Moreover, the necessity of using continuous assessment is confirmed by the clear differences between the DT- and CT-TDI flexibility, as shown in Fig. 10.

Another point of interest is the computational time required to solve these problems, as listed in Table I. Although the model is initially infinite-dimensional and is difficult to solve, by transforming the objective function, using BP spline decomposition, and using parallel solving, the final computation time is comparable to that of the DT model.

TABLE I
SOLVING TIMES FOR DIFFERENT TEST SYSTEMS

| Case | | 12-node system | 141-bus system | 533-bus system |
|---|---|---|---|---|
| Time | CT | 3.44s | 6.96s | 7.51s |
| | DT | 2.81s | 6.4s | 7.23s |

In summary, these comparisons indicate that the DT model cannot effectively and continuously capture real-time variations. In contrast, the CT model can better assess real-time PQ flexibility at the boundaries over a period within an acceptable computation time, supporting the operation of the upper-level TN while ensuring its own security.

### C. Demonstration of the Decoupled PQ Flexibility Visualization Algorithm

Using the algorithm in Section III D, we explore decoupled PQ flexibility, as shown in Fig. 11. The proposed algorithm successfully yields a decoupled PQ flexibility chart, and this result is close to the maximum decoupled PQ flexibility, which satisfies the requirement of the TSO in most cases. Additionally, the computational time of this algorithm is 2–3 seconds for this case, which is acceptable for application purposes.

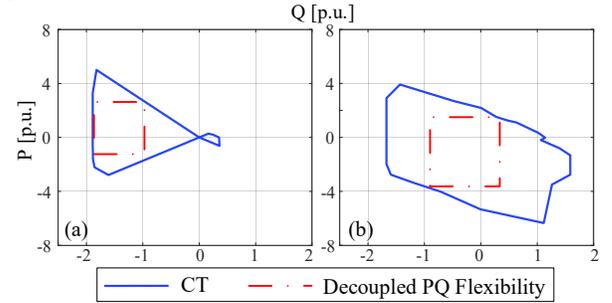

**Fig. 11.** CT-TDI flexibility and decoupled PQ flexibility of the 141-node system when (a) $t = 12$, (b) $t = 20$.

### D. Investigation of Relevant Factors

#### 1) Impact of the SOP

We mentioned that one of the features of the FDN is that the SOP becomes a viable option for adjusting network power flows. This test demonstrates the contribution of SOP to TDI flexibility. To assess the magnitude of flexibility, we will use the metric in (21), which is a weighted discrete summation of $M(t)$ introduced in (17).

$$M = \sum_{\theta_k \in \Theta} \sum_m \mathbf{1}^T \mathbf{S}_{0,B,m}, \quad (21)$$

where the relevant parameters have been previously defined. Choose $\Theta = \{0, \frac{\pi}{3}, \frac{2\pi}{3}, \pi, \frac{4\pi}{3}, \frac{5\pi}{3}\}$. Notably, we experimented with multiple sets of $\Theta$ and found that it did not change the qualitative conclusions, and therefore used one set of $\Theta$ to illustrate. If the problem is unsolvable for some $\theta_k$ values, assume $\mathbf{1}^T \mathbf{S}_{0,B,m} = 0$ under this condition.

The simulation results are listed in Table II. It can be observed that under normal operating conditions, the incorporation of SOPs enhances the CT flexibility of the FDN from 206.2 to 246.7 for 12-node system and from 0 to 196.6 for 141-bus system. The 141-bus system becomes infeasible without the SOP, that is, the value zero appears in Table II. This is primarily owing to ability of SOPs to dynamically change the topology of the FDN, manage power flows, and redistribute loads, which optimizes network performance and reduces constraints on power delivery.

In conclusion, the incorporation of an SOP enhances the flexibility of the FDN and improves its security and reliability by changing the topology of the FDN.

TABLE II
TDI FLEXIBILITY WITH AND WITHOUT THE SOP

|  | $M$ | |
|---|---|---|
|  | 12-node system | 141-bus system |
| With SOPs | 246.7 | 196.6 |
| Without SOPs | 206.2 | 0 |

*2) Impact of ESS and PV Penetration Levels*

In this section, we explore the impacts of different PV penetration levels and ESS on TDI flexibility. According to the symbol definitions in Section II, we define the PV penetration level $K_1$ and ESS capacity $K_2$ as:

$$K_1 = \frac{\sum_m \mathbf{1}^T \boldsymbol{P_{PV,B,m}}}{\sum_m \mathbf{1}^T \boldsymbol{P_{Load,B,m}}}, K_2 = \frac{4m \cdot \max\{P_D, P_C\}}{\sum_m \mathbf{1}^T \boldsymbol{P_{PV,B,m}}}. \quad (22)$$

For ease of observation, we assume the scenario described in [22] as the operational baseline, which determines $K_{10}$ and $K_{20}$, thus $K_1^* = \frac{K_1}{K_{10}}$, and $K_2^* = \frac{K_2}{K_{20}} * 100\%$.

The results are listed in Table III and Table IV.

TABLE III
TDI FLEXIBILITY WITHOUT ESS

| $K_1^*$ | $K_2^*$ | $M$ | |
|---|---|---|---|
|  |  | 12-node system | 141-bus system |
| 0 | NaN* | 41.7 | 142.9 |
| 1 | 0 | 105.3 | 196.6 |
| 2 | 0 | 140.2 | 208.8 |
| 5 | 0 | 150.0 | 211.1 |
| 10 | 0 | 155.2 | 213.7 |

*$K_2^*$ doesn't exist when $K_1^* = 0$.

TABLE IV
TDI FLEXIBILITY OF THE 12-NODE SYSTEM WITH ESS

| $K_1^*$ | $K_2^*$ | $M$ |
|---|---|---|
| 1 | 100% | 246.7 |
| 2 | 100% | 257.9 |
| 5 | 100% | 260.9 |
| 10 | 100% | 265.4 |

It can be observed from Table III that although there are uncertainties in the PV, as the penetration level of the distributed PV continues to increase from 0 to 10, the flexibility of the TDI continuously enhances from 41.7 to 155.2 or 142.9 to 213.7. However, once the penetration rate reaches a certain level, for example, $K_1^* = 5$ for a 12-node system, the growth in the TDI flexibility becomes stagnant. This is because the PV output gradually reaches the absorption limit of the FDN. At this level, the main constraint that limits TDI flexibility shifts to other system constraint, such as voltage constraint.

Additionally, from Table IV, it is evident that a reasonable configuration of the ESS can also help improve both the active and reactive power flexibilities, for example, from 105.3 to 246.7 when $K_1^* = 1$. However, it is still important to balance the installation costs and future benefits resulting from the growing flexibility of TDI.

*3) Impact of Uncertainties at Different Levels*

The risk level $\alpha$ reflects how risk averse the DSOs are, and the values of $\sigma_{PV}^2$ and $\sigma_{Load}^2$ reflect the extent uncertainty affecting the parameters. By varying the values of $\alpha$ and $\sigma_{PV}^2$, $\sigma_{Load}^2$, we observe changes of $M$ defined in (21), and the result is shown in Table V.

TABLE V
TDI FLEXIBILITY UNDER DIFFERENT UNCERTAINTIES

| $\alpha$ | $\sigma_{PV}^2$ | $\sigma_{Load}^2$ | $M$ | |
|---|---|---|---|---|
|  |  |  | 12-node system | 141-bus system |
| Deterministic | | | 262.3 | 201.2 |
| 0.01 | 0.01 | 0.01 | 170.5 | 192.7 |
| 0.05 | 0.01 | 0.01 | 176.1 | 195.3 |
| 0.1 | 0.01 | 0.01 | 246.7 | 196.6 |
|  | 0.05 | 0.01 | 167.6 | 183.6 |
|  | 0.01 | 0.05 | 183.2 | 122.4 |

It is evident from Table V that as the confidence-level $\alpha$ increases from 0.01 to 0.1, the CT flexibility of the network also increases from 170.5 to 246.7 for the 12-node system and from 192.7 to 196.6 for the 141-bus system. This is because the reduction in $\alpha$ implies an increased probability of fulfilling opportunity constraints, which tightens the feasible region of the problem by narrowing the range of conditions under which the system can operate optimally.

Similarly, increases in $\sigma_{PV}^2$ and $\sigma_{Load}^2$ from 0.01 to 0.05, which represent higher uncertainty in PV and load, respectively, could further affect the flexibility of the network by introducing more variability into the operation of the network. To ensure security operation under higher levels of uncertainty, it is necessary to tighten the feasible region, such as from 246.7 to 167.6 when $\sigma_{PV}^2$ turned to 0.05. This implies that the operational limits should be adjusted to account for greater uncertainties. Furthermore, we also notice that increases in $\sigma_{PV}^2$ and $\sigma_{Load}^2$ can lead to the problem becoming infeasible because the system cannot balance the uncertainty under the confidence-level $\alpha$ for certain $\theta_k$ values, such as $\theta_k$ values getting 167.6 and 122.4.

The results demonstrate the necessity of considering the impact of uncertainty intensity, such as variations in PV output and load, when calculating the TDI flexibility. This consideration is crucial not only for ensuring that the FDN can reliably provide the necessary flexibility to the upper-level TN, but also for guaranteeing operational security during such provisions.

V. CONCLUSION

This study introduced the novel concept of CT-TDI flexibility and demonstrated its necessities and advantages over conventional DT methods for FDNs. A more precise continuous-time mathematical model was established to capture the characteristics of SOPs in flexible distribution network (FDN) as well as the uncertainty and fluctuations of distributed PV and load, thereby providing a more accurate real-time PQ TDI assessment. Moreover, a new continuous-time metric was proposed to quantify the maximum adjustable range of active and reactive power at the TDI. An effective method was then developed to achieve more accurate CT results within a computation time comparable to that of DT evaluations. Finally, a visualization algorithm for PQ decoupling characteristics provided a decoupled PQ range for the TDI, enabling TSOs to issue dispatch instructions to the FDN based on decoupled PQ flexibility.

Case studies substantiated the advantages of the proposed CT method over the DT method. Additionally, case studies have shown that the SOP can enhance flexibility by changing the FDN topology. Studies have also validated the trends in TDI flexibility under different PV penetration rates, with or without ESS, and varying levels of uncertainty.

Overall, the proposed PQ CT-TDI flexibility calculation model and visualization method ensure that TSOs can fully utilize the potential of the FDN without compromising its security, thereby providing reliable support to the upper-level grid.

Future work will focus on establishing CT-TDI flexibility involving (1) transient constraints and considering fault conditions, and (2) leveraging a more accurate linear power flow model in modeling to obtain a more realistic TDI flexibility, particularly when the three-phase imbalance becomes prominent.

REFERENCES

[1] H. Ji et al., "Peer-to-peer electricity trading of interconnected flexible distribution networks based on distributed ledger", *IEEE Trans. Ind. Inform.*, vol. 18, no. 9, pp. 5949-5960, Sep. 2022.
[2] B. Liu, K. Meng, Z. Y. Dong, P. K. Wong and T. Ting, "Unbalance mitigation via phase-switching device and static Var compensator in low-voltage distribution network", *IEEE Trans. Power Syst.*, vol. 35, no. 6, pp. 4856-4869, Nov. 2020.
[3] Z. Liu and L. Wang, "A robust strategy for leveraging soft open points to mitigate load altering attacks", *IEEE Trans. Smart Grid*, vol. 13, no. 2, pp. 1555-1569, Mar. 2022.
[4] P. Li, Q. Wu, M. Yang, Z. Li and N. D. Hatziargyriou, "Distributed distributionally robust dispatch for integrated transmission-distribution systems", *IEEE Trans. Power Syst.*, vol. 36, no. 2, pp. 1193-1205, Mar. 2021.
[5] T. Zhang, J. Wang, H. Wang, J. Ruiyang, G. Li and M. Zhou, "On the coordination of transmission-distribution grids: A dynamic feasible region method", *IEEE Trans. Power Syst.*, vol. 38, no. 2, pp. 1857-1868, Mar. 2023.
[6] T. Chen, Y. Song, D. J. Hill and A. Y. S. Lam, "Enhancing flexibility at the transmission-distribution interface with power flow routers", *IEEE Trans. Power Syst.*, vol. 37, no. 4, pp. 2948-2960, Jul. 2022.
[7] D. M. Gonzalez et al., "Determination of the time-dependent flexibility of active distribution networks to control their TSO-DSO interconnection power flow", *Proc. IEEE 20th Power Syst. Comput. Conf.*, pp. 1-8, 2018.
[8] D. A. Contreras and K. Rudion, "Verification of linear flexibility range assessment in distribution grids", *Proc. IEEE Milan PowerTech*, pp. 1-6, 2019.
[9] M. Kalantar-Neyestanaki, F. Sossan, M. Bozorg and R. Cherkaoui, "Characterizing the reserve provision capability area of active distribution networks: A linear robust optimization method", *IEEE Trans. Smart Grid*, vol. 11, no. 3, pp. 2464-2475, May 2020.
[10] Z. Li, J. Wang, H. Sun, F. Qiu and Q. Guo, "Robust estimation of reactive power for an active distribution system", *IEEE Trans. Power Syst.*, vol. 34, no. 5, pp. 3395-3407, Sep. 2019.
[11] X. Dai, Y. Guo, Y. Jiang, C. Jones, G. Hug and V. Hagenmeyer. "Real-Time Coordination of Integrated Transmission and Distribution Systems: Flexibility Modeling and Distributed NMPC Scheduling", 2024, *arXiv:2402.00508*.
[12] M. Bolfek and T. Capuder, "An analysis of optimal power flow based formulations regarding DSO-TSO flexibility provision", *Int. J. Electr. Power Syst.*, vol. 131, Oct. 2021.
[13] J. Silva, J. Sumaili, R. J. Bessa, L. Seca, M. Matos and V. Miranda, "The challenges of estimating the impact of distributed energy resources flexibility on the TSO/DSO boundary node operating points", *Comput. Oper. Res.*, vol. 96, pp. 294-304, Aug. 2018.
[14] J. Silva et al., "Estimating the active and reactive power flexibility area at the TSO-DSO interface", *IEEE Trans. Power Syst.*, vol. 33, no. 5, pp. 4741-4750, Sep. 2018.
[15] S. Stanković, L. Söder, Z. Hagemann and C. Rehtanz, "Reactive power support adequacy at the DSO/TSO interface", *Electric Power Syst. Res.*, vol. 190, Jan. 2021.
[16] F. Capitanescu, "TSO–DSO interaction: Active distribution network power chart for TSO ancillary services provision", *Electr. Power Syst. Res.*, vol. 163, pp. 226-230, Oct. 2018.
[17] J. Silva et al., "A Data-driven Approach to Estimate the Flexibility Maps in Multiple TSO-DSO Connections", *IEEE Trans. Power Syst.*, vol. 38, no. 2, pp. 1908-1919, July 2023.
[18] Y. Liu, Z. Li and J. Zhao, "Robust data-driven linear power flow model with probability constrained worst-case errors", *IEEE Trans. Power Syst.*, vol. 37, no. 5, pp. 4113-4116, Sep. 2022.
[19] M. Parvania and A. Scaglione, "Unit commitment with continuous-time generation and ramping trajectory models", *IEEE Trans. Power Syst.*, vol. 31, no. 4, pp. 3169-3178, Jul. 2016.
[20] L. Le, J. Fang, X. Ai, S. Cui and J. Wen, "Aggregation and scheduling of multi-chiller HVAC systems in continuous-time stochastic unit commitment for flexibility enhancement", *IEEE Trans. Smart Grid*, vol. 14, no. 4, pp. 2774-2785, 2023.
[21] J. Liu et al., "Continuous-Time Aggregation of Massive Flexible HVAC Loads Considering Uncertainty for Reserve Provision in Power System Dispatch", *IEEE Trans. Smart Grid*, early access, doi: 10.1109/TSG.2024.3398627.
[22] Jun. 28, 2024. [Online]. Available: https://github.com/YYY-maker130/Assessment-of-Continuous-time-Transmission-Distribution-Interface-Active-and-Reactive-Flexibility.
[23] F. Shen et al., "Transactive Energy-Based Sequential Load Restoration of Distribution Systems with Networked Microgrids Under Uncertainty", *IEEE Trans. Smart Grid*, vol. 15, no. 3, pp. 2601-2613, May 2024.
[24] M. Baran and F. Wu, "Network reconfiguration in distribution systems for loss reduction and load balancing", *IEEE Trans. Power Del.*, vol. 4, no. 2, pp. 1401-1407, Apr. 1989.
[25] IEEE Standard for Interconnection and Interoperability of Distributed Energy Resources with Associated Electric Power Systems Interfaces, IEEE Standard 1547-2018, 2018.
[26] T. Gush and C. -H. Kim, "Robust Local Coordination Control of PV Smart Inverters with SVC and OLTC in Active Distribution Networks", *IEEE Trans. Power Del.*, vol. 39, no. 3, pp. 1610-1621, June 2024.
[27] Z. Guo, W. Wei, L. Chen, M. Shahidehpour and S. Mei, "Distribution system operation with renewables and energy storage: A linear programming based multistage robust feasibility approach", *IEEE Trans. Power Syst.*, vol. 37, no. 1, pp. 738-749, Jan. 2022.
[28] A. Bagherinezhad, M. M. Hosseini and M. Parvania, "Real-Time Hierarchical Energy Flexibility Management of Integrated Hybrid Resources", *IEEE Trans. Smart Grid*, vol. 14, no. 6, pp. 4508-4518, Nov. 2023.
[29] Narain G. Hingorani; Laszlo Gyugyi, "Static Shunt Compensators: SVC and STATCOM", in *Understanding FACTS: Concepts and Technology of Flexible AC Transmission Systems*, IEEE, 2000, pp.135-207.
[30] Y. Shen, W. Wu and S. Sun, "Stochastic Model Predictive Control Based Fast-Slow Coordination Automatic Generation Control", *IEEE Trans. Power Syst.*, vol. 39, no. 3, pp. 5259-5271, May 2024.
[31] C.-C. Liu, "A new method for the construction of maximum steady-state security regions of power systems", *IEEE Trans. Power Syst.*, vol. 1, no. 4, pp. 19-26, Nov. 1986.
[32] H. Khodr, F. Olsina, P. D. O.-D. Jesus and J. Yusta, "Maximum savings approach for location and sizing of capacitors in distribution systems", *Elect. Power Syst. Res.*, vol. 78, no. 7, pp. 1192-1203, Jul. 2008.
[33] G. Malmer and L. Thorin, *Network reconfiguration for renewable generation maximization: Application of a power-flow optimization algorithm on a distribution network in southern Sweden*. Lund University, Aug. 2023, Accessed on: Jun. 12, 2024, [Online]. Available: https://portal.research.lu.se/en/publications/network-reconfiguration-for-renewable-generation-maximization-app.
[34] A. P. Singh, V. Jain, S. Chaudhari, F. A. Kraemer, S. Werner and V. Garg, "Machine learning-based occupancy estimation using multivariate sensor nodes", *Proc. IEEE Globecom Workshops (GC Wkshps)*, pp. 1-6, Dec. 2018.